# Twisted nonlinear optics in monolayer van der Waals crystals

Tenzin Norden[1,*], Luis M. Martinez[1], Nehan Tarefder[1], Kevin W. C. Kwock[2], Luke M. McClintock[1], Nicholas Olsen[3], Luke N. Holtzman[4], Xiaoyang Zhu[3], James C. Hone[5], Jinkyoung Yoo[1], Jian-Xin Zhu[1,6], P. James Schuck[5], Antoinette J. Taylor[1], Rohit P. Prasankumar[1,7], Wilton J. M. Kort-Kamp[6,*], Prashant Padmanabhan[1,*]

**Affiliations:**

[1]Center for Integrated Nanotechnologies, Los Alamos National Laboratory; Los Alamos, NM, 87545, USA

[2]Department of Electrical Engineering, Columbia University; New York, NY, 10027, USA

[3]Department of Chemistry, Columbia University; New York, NY, 10027, USA

[4]Department of Applied Physics and Applied Mathematics, Columbia University; New York, NY 10027, USA

[5]Department of Mechanical Engineering, Columbia University; New York, NY, 10027, USA

[6]Theoretical Division, Los Alamos National Laboratory; Los Alamos, NM, 87545, USA

[7]Deep Science Fund, Intellectual Ventures; Bellevue, WA, 98005, USA

*Corresponding authors. Email: tnorden@lanl.gov, kortkamp@lanl.gov, prashpad@lanl.gov

**Abstract:** In addition to a plethora of emergent phenomena[1–3], the spatial topology of optical vortices enables an array of applications spanning communications[4] to quantum photonics[5]. Nonlinear optics is essential in this context, providing access to an infinitely large set of quantum states associated with the orbital angular momentum of light. Nevertheless, the realization of such processes have failed to keep pace with the ever-growing need to shrink the fundamental length-scale of photonic technologies to the nanometer regime[6]. Here, we push the boundaries of vortex nonlinear optics to the ultimate limits of material dimensionality. By exploiting second and third-order frequency-mixing processes in semiconducting monolayers, we demonstrate the independent manipulation of the wavelength, orbital angular momentum, and spatial distribution of vortex light-fields. Due to the atomically-thin nature of the host quantum material, this control spans a broad spectral bandwidth in a highly-integrable platform, unconstrained by the traditional limits of bulk nonlinear optical materials. Our work heralds a new avenue for ultra-compact and scalable hybrid nanotechnologies empowered by twisted nonlinear light-matter interactions in van der Waals quantum nanomaterials.

Light carries energy and momentum, the latter comprised of both linear and angular components. Circularly polarized light possesses non-zero spin angular momentum (SAM), a property that has been exploited to study a wide array of material phenomena, including valley polarization[7,8], magnetism[9–11], and topology[12,13]. Nevertheless, SAM is intrinsically restricted to a two-parameter space defined by the handedness of the light-field's polarization. In contrast, spatially-structured vortex beams, possessing helical or "twisted" wavefronts, can carry non-zero orbital angular momentum (OAM), equivalent to integer multiples of the elementary unit $\hbar$[14–16]. OAM is therefore associated with an unbounded number of orthogonal states, indexed by the degree of wavefront twisting through a parameter known as the topological charge ($\ell$). In addition to potentially mediating complex light-matter interactions[1,17–19], vortex beams are highly advantageous for applications that can leverage their infinite-dimensional space, including multiplexed optical communications[20–23] and robust quantum communication paradigms[5,24–26].

Most efforts aimed at exploiting vortex beams are predicated on an ability to precisely tune the wavelength and OAM of a light-field[27]. Nonlinear optics offers a powerful technique for manipulating both of these parameters via frequency mixing processes[28–31]. Yet, in most cases, these approaches rely on birefringent phase matching in bulk nonlinear optical (NLO) crystals[32], resulting in operational bandwidths that are tightly constrained by material choice and geometry. More importantly, bulk NLO materials are inherently ill-suited to chip-level or nanoscale platforms, limiting their use in integrated devices. These challenges have fueled an interest in uncovering NLO processes in van der Waals (vdW) materials[33–36], which possess giant nonlinear susceptibilities and are free of adverse volumetric effects at the few atomic layer length scale. Moreover, they are compatible with complementary metal–oxide–semiconductor processes[37] and allow for bond-free integration with hybrid photonic systems, dramatically enhancing their functionalities[38,39]. Nevertheless, efforts aimed at realizing broadband optical wavelength tuning via two-dimensional (2D) vdW materials have entirely focused on Gaussian beams where $\ell = 0$, leaving untapped their potential as nanoscale sources of tunable vortex light. The incorporation of the OAM degree of freedom into nanoscale nonlinear optics would therefore open a path to dramatically expand the utility and shrink the length scale of a vast array of classical and quantum communication technologies.

Here, we show that vdW quantum materials provide a transformative pathway for on-demand, nanoscale control of the fundamental properties of optical vortices via second- and third-order NLO processes[40]. Using monolayer transition metal dichalcogenides (TMDs), we demonstrate independent control over the wavelength, topological charge, and radial profile of twisted light-fields through difference frequency generation (DFG), sum frequency generation (SFG), and four-wave mixing (FWM) (Fig. 1a). As these nonlinear phenomena are supported in a single atomic layer, they are free from dispersion-induced phase mismatch, allowing for broadband operation that further benefits from the intrinsically large optical nonlinearities of monolayer TMDs. Taken together, our results demonstrate the potential and versatility of vdW crystals as a material platform for highly tunable vortex light at the nanoscale.

**Manipulating vortex pulses through difference frequency mixing in a vdW monolayer**
DFG is a second-order nonlinear frequency down-conversion process that is integral to optical parametric amplification (OPA)[41], which forms the basis of modern tunable coherent light sources. Here, the interaction of a high energy pump photon and a low energy seed photon leads to the generation of a photon at their energy difference (i.e., $\hbar\omega_{DFG} = \hbar\omega_p - \hbar\omega_s$, left panel of Fig. 1b). To investigate vortex DFG in a single-layer TMD, we utilized a Gaussian pump ($\hbar\omega_p$ =3.10 eV, $\ell_p = 0$) and a Laguerre-Gaussian vortex seed ($\ell_s \neq 0$). Fig. 1c shows the spectrum of the DFG output from a monolayer of the canonical semiconducting TMD MoS₂ for various seed photon energies ($\hbar\omega_s = 1.88 - 1.96$ eV) at different values of $\ell_s$. It is immediately evident that the vdW crystal supports broad spectrum frequency conversion[42]. More importantly, the spectrum of the DFG output is insensitive to the value of $\ell_s$, providing compelling evidence that wavelength and topological charge conversion are decoupled processes.

The polarization of frequency mixed outputs resulting from NLO processes are uniquely associated with the point group symmetry of the crystal. Therefore, we performed polarimetry analysis to confirm that the DFG output originated from the MoS₂ layer. Fig. 1d shows the polarization dependence of the DFG output intensity with respect to the relative angle of the pump and seed polarization when $\ell_s = 1$. The polarization angle of the DFG output ($\theta_{DFG}$) was measured by rotating an analyzer placed before the detector. As shown in the top panel schematic of Fig. 1d, the seed polarization angle was fixed to the crystal's armchair direction ($\theta_s = 0°$), and the pump polarization was either collinear ($\theta_p = 0°$, blue pattern in the bottom panel of Fig. 1d) or orthogonal ($\theta_p = 90°$, red pattern in the bottom panel of Fig. 1d) to this axis. The bi-lobed structure and the $\pi/2$ shift in the orientations of the two patterns match the theoretically expected polarization direction $\theta_{DFG} = \pi/2 - \theta_s - \theta_p$ for the DFG signal due to monolayer MoS₂, given its D$_{3h}$ point group symmetry (see Supplementary section VII). Additionally, the intensity of the DFG output measured without an analyzer does not show any dependence on the relative polarization angles of the pump and the seed[42] (bottom panel of Fig. 1e). Similar analysis for seed pulses with $\ell_s = 0 - 3$ show identical results for both MoS₂ and WSe₂ monolayers (see Supplementary Figs. S3 and S5). This indicates the lack of appreciable impact of the seed's topological charge on the polarization properties of the DFG output, highlighting their relative independence in the vortex DFG process.

An analysis of the time-domain dynamics of the DFG output intensity, relative to the delay between the pump and seed pulses reveals a Gaussian temporal profile (blue trace in Fig. 1f) associated with a time-dependent buildup of the DFG spectrum (Fig. 1g). This implies that the DFG process occurs only when the seed and pump pulses are spatiotemporally overlapped. In contrast, the reflected seed shows no appreciable change in either intensity (red trace in Fig. 1f) or spectrum (Fig. 1h) over the timeframe of the DFG process. The absence of any seed enhancement, consistent with previous reports for Gaussian beams[42], is a clear indication that parametric amplification is either absent or exceptionally weak. This is likely due to the low pump-to-seed intensity ratio in our experiment[41], stemming from the need to mitigate sample damage while ensuring a seed with an intensity above our detection threshold and spectral range within the operational regime of the spatial light modulator (SLM).

We directly imaged the seed and DFG beams to characterize their OAM state. Fig. 2a shows the real-space seed beam images for $\ell_s = 0 - 6$, revealing a characteristic annular intensity profile for $\ell_s \neq 0$ and diameter that is correlated to the magnitude of the topological charge (i.e., a larger $\ell_s$ leads to a seed with a larger geometric radius). Momentum space mapping allows for the quantification of the topological charge via an OAM-dependent fringing and skew of the beam at the focal plane of a cylindrical lens[43]. Here, the number of fringes in the seed, $N_F^S$, is equivalent to $\ell_s + 1$, as confirmed in the maps shown in Fig. 2b. Intriguingly, the DFG output inherits the annular intensity profile, with its diameter being approximately equivalent to that of the seed for any given value of $\ell_s$ (Fig. 2d). Moreover, the momentum space mapping reveals the presence of fringes with $N_F^{DFG} = N_F^S$ but with skew directions that are opposite that of the seed (Fig. 2c). Taken together, this indicates a conservation of the magnitude but an inversion of the sign of the topological charge in the vortex DFG process.

To contextualize our observations, we theoretically consider the case of a monolayer NLO crystal on an inversion symmetric substrate illuminated by a structured optical field. The incident light is modeled as the superposition of $N$ monochromatic waves with frequencies $\{\omega_1, \omega_2, \cdots \omega_N\}$. This allows for the derivation of the reflected frequency-mixed electric field as

$$\boldsymbol{E}_R(\boldsymbol{R},\omega) = 2\pi\hbar \sum_{m=1} \sum_{\substack{(\zeta_1,\zeta_2\cdots\zeta_m) \\ \in \{\pm\omega_j\}}} \mathbb{R}^{(m)}(\{\zeta_j\}) : \boldsymbol{\mathcal{E}}_0^{\zeta_1}(\boldsymbol{R}) \cdots \boldsymbol{\mathcal{E}}_0^{\zeta_m}(\boldsymbol{R}) \delta\left(\hbar\omega - \sum_{j=1}^{m} \hbar\zeta_j\right), \quad (1)$$

where $m$ is the harmonic order, $\mathbb{R}^{(m)}$ is the generalized Fresnel reflection coefficient tensor, and $\boldsymbol{\mathcal{E}}_0^{\zeta}$ is the vectorial spatial profile of the incident field oscillating at frequency $\zeta \in \{\pm\omega_1, \pm\omega_1, \ldots, \pm\omega_N\}$. This expression is valid for arbitrary nonlinear orders and for any structured light fields within the paraxial approximation, while explicitly accounting for the 2D nature of the quantum material (see Methods).

Equation (1) allows us to draw several conclusions regarding vortex NLO processes involving 2D materials. First, for the case where there are only two incident fields (i.e., pump and seed) with energy $\hbar\omega_p$ and $\hbar\omega_s$ (i.e., $\zeta_j = \{\pm\omega_p, \pm\omega_s\}$), the Dirac delta in Eq. (1) enforces energy conservation,

$$\hbar\omega_{\text{out}} = \hbar|\alpha\,\omega_p + \beta\,\omega_s|. \quad (2)$$

Here, $-m \leq \alpha, \beta \leq m$ are integers corresponding to the difference between the number of positive and negative $\omega_p$ and $\omega_s$ field components contributing to the nonlinear process. For the case of DFG, Eq. (2) reduces to the observed $\hbar\omega_{\text{out}} = \hbar|\omega_p - \omega_s|$. Moreover, Eq. (1) does not impose any conditions over the linear momentum of the input or reflected fields. This implies that 2D vdW crystals enable broadband frequency conversion for all multibeam frequency mixing processes, regardless of the input fields' spatial profile or OAM, in agreement with the OAM-agnostic DFG tuning bandwidth seen in Fig. 1c. In addition, material properties and symmetries are entirely encoded in $\mathbb{R}^{(m)}$, allowing them to modulate the light-fields' polarization degrees of freedom (and therefore SAM), but not their spatial structure or OAM, as seen from the invariance of the DFG polarimetry results for various values of $\ell_s$ (see Supplementary Fig. S3). This asymmetry between the coupling of material degrees of freedom to OAM and SAM is also at the heart of the complexity in driving electronic currents and transferring OAM from weakly focused beams[1] to charge carriers in solids. Lastly, when the incident fields are eigenmodes of the paraxial wave equation with well-defined topological charge $\ell_j$ (i.e., $\boldsymbol{\mathcal{E}}_0^{\zeta_j} \propto e^{\pm i l_j \phi}$), Eq. (1) implies that the OAM of the output fields, due to an $m^{\text{th}}$ order nonlinear process, is determined by the sum of the OAM of all contributing input beams. For the case of a pump and seed with topological charge $\ell_p$ and $\ell_s$, respectively,

$$\ell_{\text{out}} = \alpha\,\ell_p + \beta\ell_s. \quad (3)$$

It is evident that the frequency-mixed field inherits the OAM of the constituent inputs (e.g., $\alpha, \beta = +1, -1$ for DFG), consistent with the results presented in Fig. 2.

**Sum frequency generation and radial mode matching**

SFG is the second-order counterpart to DFG and involves the frequency up-conversion of two incident photons, with potentially different photon energies, producing a photon with energy $\hbar\omega_{SFG} = \hbar\omega_p + \hbar\omega_s$ (middle panel of Fig. 1b). To demonstrate the vortex SFG process, we performed a similar experiment on a monolayer of MoS$_2$, utilizing a Gaussian pump ($\hbar\omega_p$ =1.63 eV, $\ell_p = 0$, Fig. 3a) and a Laguerre-Gaussian vortex seed ($\hbar\omega_s$=1.18 eV, $\ell_s = +3$, Fig. 3b). As seen in Fig. 3c, the SFG output again inherits the annular intensity profile of the seed. However, in contrast to the DFG case, though $N_F^{SFG} = N_F^S$, the seed and SFG momentum-space maps are skewed in the same direction. This confirms the equivalence of both the magnitude and sign of the topological charge of the two beams, which can be understood from Eq. (3), with $\alpha = \beta = +1$. Similar results were obtained for the other values of $\ell_s$ (see Supplementary Fig. S6), in addition to the broad wavelength tunability regardless of the seed's topological charge (see Supplementary Fig. S7).

Apart from topological charge, Laguerre-Gaussian vortices have another indexed spatial degree of freedom, namely the radial index $n$; a vortex beam with $n \neq 0$ is characterized by an intensity profile comprised of $n + 1$ concentric annuli. We have exclusively considered seed pulses with $n_s = 0$ thus far, but now turn to processes where $n_s \neq 0$. Radial mode conversion in a two-beam second-order mixing process is determined by an overlap integral between the pump, seed, and frequency-mixed fields involved in the nonlinear process, given by (see Methods)

$$\Lambda^{n_p n_s n_i}_{\ell_p \ell_s \ell_i} = 2\pi\, \delta_{\ell_i, \ell_p \pm \ell_s} \int_0^\infty R\, u^{\omega_p}_{n_p,\ell_p}(R) u^{\omega_s}_{n_s,\ell_s}(R) u^{*}_{n_i,\ell_i}(R)\, dR, \quad (4)$$

where $u^\omega_{n,\ell}$ ($= u_{n,\ell}, \omega \geq 0$; $u^*_{n,\ell}, \omega < 0$) are the orthonormal Laguerre-Gaussian modes and the $\pm$ symbol appearing in the subscript of the Kronecker delta function enforces OAM conservation for SFG and DFG, respectively. We consider the case where the pump is a Gaussian mode (i.e., $n_p = \ell_p = 0$) focused on the monolayer. When the seed is a Laguerre-Gaussian vortex also focused on the monolayer, the mode functions $u_{n,\ell}$ are real-valued and depend on the absolute value of $\ell$. As a result, the $\Lambda$ values are identical for SFG and DFG, allowing us to concentrate on the former without loss of generality. We begin with the case where the pump beam waist (~22 µm) is significantly larger than that of the Laguerre-Gaussian vortex seed (~3 µm, see Fig. 3d). The plot of $\Lambda$ in Fig. 3e shows a nearly perfect radial mode matching in the SFG process, an effective conservation law in which $n_{SFG} = n_s$. This can be clearly seen in simulations of the SFG output intensity profile for a seed with $\ell_s = 1$ and $n_s = 1 - 4$ (top panels of Fig. 3f) and is in remarkably close agreement with the experimentally observed SFG output profiles from a monolayer of MoS$_2$ (bottom panels of Fig. 3f).

A considerably different situation occurs in a geometry where the Gaussian pump beam waist (~4 µm) is comparable to the seed. As shown in Fig. 3g, when the seed is then imparted with non-zero topological charge, the pump beam is overlapped with only the central annulus of the seed. As a result, mode conservation breaks down, giving way to a distribution of radial modes in the SFG output for any given $n_s$ (Fig. 3h). The result is a single diffuse annulus in the far-field, which can be seen in both the simulated (top panel of Fig. 3i) and experimentally obtained (bottom panel of Fig. 3i) SFG output profiles.

**Four-wave mixing**
Thus far, we have discussed three-wave vortex NLO processes enabled by the second-order nonlinearity of monolayer crystals. However, FWM is also possible through the third-order susceptibility[40], $\chi^{(3)}$. Here, two pump photons and a seed photon mix to yield a FWM output with photon energy $\hbar\omega_{FWM} = \hbar\omega_{p1} + \hbar\omega_{p2} - \hbar\omega_s$ (right panel of Fig. 1b). FWM is important in applications such as extreme ultraviolet light generation and control[44,45], near-field imaging[46], frequency comb generation[47], and quantum state generation[48], making its nanoscale realization with vortex light of particular interest. In our experiment, we make use of two-photon excitation by the same Gaussian pump, and as such, $\hbar\omega_{p1} = \hbar\omega_{p2} = \hbar\omega_p = 1.54$ eV. As shown in Fig. 4a, the MoS$_2$ monolayer supports vortex FWM over a range of seed photon energies, with nearly equivalent spectral profiles regardless of the value of $\ell_s$.

The FWM process is driven by a nonlinear polarization source term of the form $P_i^{NL} \propto \chi_{ijkl}^{(3)} E_j^p E_k^p E_l^s$, where $E^p$ ($E^s$) is the electric field of the pump (seed) and $i, j, k$, and $l$ are Cartesian directions. This leads to an expected square-law dependence on the pump power and linear dependence on seed power, which we observe experimentally as shown in Figs. 4b and 4c, respectively. The slight drop in the intensity of the FWM output for the $\ell_s = 2$ case is most likely due to the decreased efficiency of the FWM process for seeds with larger geometric diameters, owing to their diminished seed intensity and mode overlap with the Gaussian pump. In addition, considering the form of $P_i^{NL}$ for FWM, the intensity angular distribution is expected to show a bi-lobed pattern oriented along the seed polarization direction (see Supplementary section VII), which we confirmed experimentally as shown in the right-panel of Fig. 4d. Finally, the top right panel of Fig. 4e shows the annular intensity pattern of the FWM signal, qualitatively similar in size to the seed pulse (top left panel, Fig. 4e). However, the sign of the topological charge of the FWM signal is inverted with respect to the seed, as seen from the opposing skews of the momentum space maps of the seed (lower-left panel, Fig. 4e) and FWM output (lower-right panel, Fig. 4e). In analogy to the DFG and SFG processes, the OAM conservation law for FWM obtained from Eq. (3) is $\ell_{FWM} = 2\ell_p - \ell_s$. Given that $\ell_p = 0$, we find that $\ell_{FWM} = -\ell_s = -3$ as observed experimentally. While we have discussed the case for $\ell_s = 3$, results for other values of $\ell_s$ are consistent with this analysis (see Supplementary Fig. S8). Ultimately, this confirms that wavelength tunability and topological charge conversion are fully decoupled, even in higher order vortex NLO processes, confirming the higher-harmonic validity of Eq. (1).

**Discussion**
Our results underscore the ability of monolayer vdW materials to enable independent control of the photon energy, topological charge, and radial mode of optical vortices through second- and third-order frequency-mixing processes. Their ability to support such structured NLO processes without phase-matching constraints and with bond-free compatibility with nanophotonic devices[49–51] implies that these atomically-thin materials may provide a compelling route to realize monolithic nanoscale sources of broadly tunable vortex and higher-order structured light. This, in turn, could lead to revolutionary advancements in our ability to develop integrated nanodevice platforms for a multitude of applications, including high-density optical data transmission[20–23], super-resolution imaging[52,53], and quantum information[5,24–26]. Moreover, while we have focused on DFG, SFG, and FWM, we envision that vdW monolayers can also support other exotic NLO phenomena. This could potentially push the boundary of nanoscale vortex nonlinear optics to the extreme ultraviolet regime through high-harmonic generation or enable the creation of tunable twisted quantum states of light through processes such as spontaneous parametric down-conversion. We anticipate that many new opportunities will emerge at the intersection of structured light and vdW quantum nanomaterials through the exploitation of spatio-temporal light-matter interactions.

**Methods**

***Crystal synthesis and sample preparation.*** We utilized commercial $MoS_2$ (HQ Graphene), monolayer $MoS_2$ films grown on $SiO_2$/Si substrates via chemical vapor deposition, and high quality single-crystals of $WSe_2$ synthesized using a previously described self-flux method[54]. Large-area monolayers of $MoS_2$ were exfoliated via the Au tape exfoliation method[55]. Here, a 150 nm thin layer of Au tape was prepared by depositing Au onto a polished silicon wafer, followed by spin coating with a protective layer of polyvinylpyrrolidone (PVP). The Au tape was then removed from the silicon wafer using thermal release tape (Semiconductor Equipment Corp. Revalpha RA-95LS(N)). The large-area monolayer flake was mechanically exfoliated by lightly pressing the Au tape onto the surface of the bulk $MoS_2$. The Au tape was placed onto a 0.3 mm glass substrate, which was then heated on a hot plate at 135 °C to remove the thermal release tape. Once the thermal release tape is removed, the PVP protection layer was dissolved by soaking the substrate in deionized water for 3 hours and acetone for 1 hour. Subsequently, the Au film was dissolved in a $I_2$/KI etchant solution for 5 min. Finally, the sample was again soaked in deionized water for 2 hours, before rinsing it with isopropanol and drying with $N_2$ gas. The single layer character of the sample was confirmed using optical contrast and Raman spectroscopy (Supplementary Figs. S1a. and S1b). Small-area monolayer $WSe_2$ flakes were prepared using standard mechanical exfoliation techniques.

***Time-resolved vortex NLO spectroscopy and imaging.*** NLO experiments were conducted on a time-resolved structured light microscopy system (see Supplementary Fig. S1c). A tunable Ti:sapphire oscillator (1.15 – 1.82 eV, ~150 fs, ~80 MHz) was used to pump an optical parametric oscillator (OPO), which emitted a tunable signal in the near-infrared (0.78 – 1.24 eV, ~150 fs, ~80 MHz). In all our studies, the pump pulse supplied by the oscillator was kept as a Gaussian ($\ell_p = 0$) and the seed pulse supplied by the OPO was spatially structured into a Laguerre-Gaussian optical vortex ($\ell_s \neq 0$) using a liquid crystal on silicon phase-only SLM with an operational range of 1.13 – 2.95 eV. Here, the Gaussian seed beam from the OPO was first expanded with a telescope, and then converted into a vortex beam after being reflected from the SLM, which was encoded with a particular phase mask (see Supplementary Fig. S8), at near-normal incidence. The pump (Gaussian) and the seed (vortex) pulses were then combined collinearly using a 950 nm short-pass dichroic filter, reflected with a 50:50 non-polarizing beam splitter, and focused onto the sample plane using a 20X (0.42 NA) infinity-corrected apochromatic objective. The pulses were spatiotemporally overlapped on the crystal by adjusting a mechanical delay line that was part of the pump beam path. The focal spot diameter of the seed on the sample ranged from 4 – 12 μm for $\ell_s = 0 - 6$, respectively. Therefore, to ensure full spatial overlap of the pump and seed on the sample, the pump's focal spot diameter was set to ~ 15 μm (except for the radial mode matching studies, as described above).

For the SFG and FWM processes, the photon energy of the pump was set to 1.63 eV and 1.54 eV, respectively, while for the DFG process, the pump was frequency doubled to 3.10 eV using a 1 mm thick type-I bismuth borate crystal. The DFG, SFG and FWM outputs were collected through the same objective in a reflection geometry and picked off using a 490 nm short pass dichroic, 650 nm long pass dichroic, or a 50:50 600-1700 nm beam splitter, respectively. The pump and seed beams were blocked by placing a combination of interference filters in the collection path before the detectors to isolate the NLO output of interest. A spectrometer equipped with a thermoelectrically cooled charge-coupled device (CCD) camera was used to record the spectrum of the generated outputs. For the polarization and intensity dependent measurements, the output beam was directed into a photomultiplier tube (PMT). Here the pump beam was modulated with an optical chopper and the PMT signal was fed to a lock-in amplifier referenced to the chopper frequency. The incident pump and signal polarization angles were carefully set independently with reference to the crystallographic armchair axis of the crystal using $\lambda/2$ waveplates, and the polarization dependence of all outputs were measured by rotating an analyzer placed before the PMT. The analyzer was also removed to measure the raw intensity of the NLO outputs under pump polarization rotation by rotating the pump's $\lambda/2$ waveplate. All beams were also directed to a silicon electron-multiplying CCD (EMCCD) camera, to either image their intensity profiles or map them to momentum space using a cylindrical lens (f = 120 mm), placed before the EMCCD, to determine the magnitude and sign of their topological charge.

***Theory of nonlinear vortex light scattering in 2D van der Waals crystals.*** We consider the case of a monolayer crystal lying over a substrate with linear refractive index $n(\omega)$ that is illuminated by monochromatic structured light fields. The electromagnetic fields on the monolayer surface ($z = 0$) at a position $\boldsymbol{R}$ and with frequency $\omega$ satisfy[56]

$$\hat{\boldsymbol{z}} \times [\boldsymbol{E}_T(\boldsymbol{R}, \omega) - \boldsymbol{E}_0(\boldsymbol{R}, \omega) - \boldsymbol{E}_R(\boldsymbol{R}, \omega)] = 0,$$

$$\hat{\boldsymbol{z}} \times [\boldsymbol{H}_T(\boldsymbol{R}, \omega) - \boldsymbol{H}_0(\boldsymbol{R}, \omega) - \boldsymbol{H}_R(\boldsymbol{R}, \omega)] = \boldsymbol{J}(\boldsymbol{R}, \omega),$$

where $\boldsymbol{J}(\boldsymbol{R}, \omega)$ is the monolayer's induced surface current density due to both linear and nonlinear frequency mixing processes. Also, $\boldsymbol{E}_{0,R,T}, \boldsymbol{H}_{0,R,T}$ are the incident, reflected, and transmitted electromagnetic fields, respectively. In the following, we assume that the input beams impinge normal to the monolayer, neglect nonlinear effects due to the substrate, and consider only contributions up to leading order in the paraxial approximation (i.e., $\boldsymbol{H}_{0,R}(\boldsymbol{R}, \omega) \simeq \pm \frac{1}{\mu_0 c} \hat{\boldsymbol{z}} \times \boldsymbol{E}_{0,R}(\boldsymbol{R}, \omega)$ and $\boldsymbol{H}_T(\boldsymbol{R}, \omega) \simeq \frac{n(\omega)}{\mu_0 c} \hat{\boldsymbol{z}} \times \boldsymbol{E}_T(\boldsymbol{R}, \omega)$). Hence,

$$\boldsymbol{E}_T(\boldsymbol{R}, \omega) = \frac{2}{1+n(\omega)} \boldsymbol{E}_0(\boldsymbol{R}, \omega) - \frac{\mu_0 c}{1+n(\omega)} \boldsymbol{J}(\boldsymbol{R}, \omega),$$

$$\boldsymbol{E}_R(\boldsymbol{R}, \omega) = \frac{1-n(\omega)}{1+n(\omega)} \boldsymbol{E}_0(\boldsymbol{R}, \omega) - \frac{\mu_0 c}{1+n(\omega)} \boldsymbol{J}(\boldsymbol{R}, \omega).$$

To compute $\boldsymbol{J}(\boldsymbol{R}, \omega)$, we note that in the weak interaction regime the time-domain surface current density can be expanded in powers of the electromagnetic field on the surface of the monolayer as $\boldsymbol{J}(\boldsymbol{R}, t) = \sum_m \boldsymbol{J}^{(m)}(\boldsymbol{R}, t)$, where[57]

$$\boldsymbol{J}^{(m)}(\boldsymbol{R}, t) = \frac{1}{(2\pi)^{3m}} \int d\boldsymbol{q}_1 \cdots d\boldsymbol{q}_m \int d\omega_1' \cdots d\omega_m' e^{i \sum_{j=1}^m (\boldsymbol{q}_j \cdot \boldsymbol{R} - \omega_j' t)} \vec{\sigma}^{(m)}(\{\boldsymbol{q}_j\}; \{\omega_j'\}) : \boldsymbol{E}_T(\boldsymbol{q}_1, \omega_1') \cdots \boldsymbol{E}_T(\boldsymbol{q}_m, \omega_m')$$

is the $m^{\text{th}}$ order nonlinear contribution to the surface electronic current. The notation $\{\boldsymbol{q}_j\}, \{\omega_j'\}$ denotes the entire set of linear momentum and frequency variables that the nonlinear conductivity tensor $\overleftrightarrow{\sigma}^{(m)}$ depends upon. In practice, however, the linear momentum per photon for propagative light fields is significantly smaller than that of the charge carriers in the monolayer. Thus, one can neglect spatial dispersion in the nonlinear conductivity, resulting in

$$\boldsymbol{J}^{(m)}(\boldsymbol{R},t) = \frac{1}{(2\pi)^m}\int d\omega_1' \cdots d\omega_m' e^{-i\sum_{j=1}^m \omega_j' t} \overleftrightarrow{\sigma}^{(m)}(\{\omega_j'\}) : \boldsymbol{E}_T(\boldsymbol{R},\omega_1') \cdots \boldsymbol{E}_T(\boldsymbol{R},\omega_m').$$

Fourier-transforming the previous equation to the frequency space,

$$\boldsymbol{J}(\boldsymbol{R},\omega) = \sum_m \frac{1}{(2\pi)^{m-1}}\int d\omega_1' \cdots d\omega_m' \overleftrightarrow{\sigma}^{(m)}(\{\omega_j'\}) : \boldsymbol{E}_T(\boldsymbol{R},\omega_1') \cdots \boldsymbol{E}_T(\boldsymbol{R},\omega_m') \delta(\omega - \sum_{j=1}^m \omega_j').$$

When computing the fields, it is convenient to explicitly write the linear conductivity contribution to the current and use the fact that the symmetry group ($D_{3h}$) of monolayer MoS2 enforces that second-rank tensors are isotropic, i.e., $\overleftrightarrow{\sigma}^{(1)}(\omega_j) = \sigma^{(1)}(\omega_j)\mathbb{I}$, where $\mathbb{I}$ is the identity operator. The incident field due to a superposition of $N$ monochromatic waves is $\boldsymbol{E}_0(\boldsymbol{R},\omega) = 2\pi \sum_{\zeta_0 \in \{\pm\omega_j\}} \boldsymbol{\mathcal{E}}_0^{\zeta_0}(\boldsymbol{R})\delta(\omega - \zeta_0)$, where the frequency superscript in $\boldsymbol{\mathcal{E}}_0^{\omega_j}(\boldsymbol{R})$ indicates that we should take the complex conjugate for negative frequency components of the field. Hence, the transmitted and reflected fields up to leading order in $\overleftrightarrow{\sigma}^{(m)}$ are given by

$$\boldsymbol{E}_T(\boldsymbol{R},\omega) = 2\pi\hbar \sum_{m=1} \sum_{\substack{(\zeta_1,\zeta_2\cdots\zeta_m) \\ \in \{\pm\omega_j\}}} \mathbb{T}^{(m)}(\{\zeta_j\}) : \boldsymbol{\mathcal{E}}_0^{\zeta_1}(\boldsymbol{R}) \cdots \boldsymbol{\mathcal{E}}_0^{\zeta_m}(\boldsymbol{R}) \delta(\hbar\omega - \sum_{j=1}^m \hbar\zeta_j),$$

$$\boldsymbol{E}_R(\boldsymbol{R},\omega) = 2\pi\hbar \sum_{m=1} \sum_{\substack{(\zeta_1,\zeta_2\cdots\zeta_m) \\ \in \{\pm\omega_j\}}} \mathbb{R}^{(m)}(\{\zeta_j\}) : \boldsymbol{\mathcal{E}}_0^{\zeta_1}(\boldsymbol{R}) \cdots \boldsymbol{\mathcal{E}}_0^{\zeta_m}(\boldsymbol{R}) \delta(\hbar\omega - \sum_{j=1}^m \hbar\zeta_j),$$

where the $m^\text{th}$ order generalized nonlinear Fresnel transmission and reflection coefficient tensors are

$$\mathbb{T}^{(m)}(\{\zeta_j\}) = \delta_{m,1}\mathcal{T}(\zeta_1)\mathbb{I} - (1 - \delta_{m,1})\frac{\mu_0 c\,\mathcal{T}(\zeta_1) \times \mathcal{T}(\zeta_2) \cdots \times \mathcal{T}(\zeta_m)}{1 + n(\omega) + \mu_0 c\,\sigma^{(1)}(\omega)} \overleftrightarrow{\sigma}^{(m)}(\{\zeta_j\}),$$

$$\mathbb{R}^{(m)}(\{\zeta_j\}) = \delta_{m,1}\mathcal{R}(\zeta_1)\mathbb{I} - (1 - \delta_{m,1})\frac{\mu_0 c\,\mathcal{T}(\zeta_1) \times \mathcal{T}(\zeta_2) \cdots \times \mathcal{T}(\zeta_m)}{1 + n(\omega) + \mu_0 c\,\sigma^{(1)}(\omega)} \overleftrightarrow{\sigma}^{(m)}(\{\zeta_j\})$$

where $\omega = \sum_{j=1}^m \zeta_j$ and $\mathcal{T}(\omega) = \frac{2}{1+n(\omega)+\mu_0 c\,\sigma^{(1)}(\omega)}$ and $\mathcal{R}(\omega) = \frac{1-n(\omega)-\mu_0 c\,\sigma^{(1)}(\omega)}{1+n(\omega)+\mu_0 c\,\sigma^{(1)}(\omega)}$ are the corresponding the linear order coefficients.

When the incident light field are pure eigenmodes of the paraxial wave equation, they can be written as $\boldsymbol{\mathcal{E}}_0^{\omega_j}(\boldsymbol{R}) = \boldsymbol{\mathcal{A}}_0^{n_j,\ell_j} u_{n_j,\ell_j}(R)e^{i\,\ell_j\varphi}$, where $\boldsymbol{\mathcal{A}}_0^{n_j,\ell_j}$ is the field amplitude and $n_j$ is the radial index of the orthonormal functions $u_{n_j,\ell_j}(R)$. The output electromagnetic fields can be decomposed in an eigenmode superposition as $\boldsymbol{E}_{R,T}(\boldsymbol{R},\omega) = \sum_{n,\ell} \boldsymbol{\mathcal{A}}_{R,T}^{n,\ell} u_{n,\ell}(R)e^{i\,\ell\varphi}$. This results in coupling between the pumps/seeds eigenmodes and conversion/generation of idler output fields with a complex spatial profile. The coupling constant for an $m^\text{th}$ order process is

$$\Lambda_{\substack{n_1\cdots n_m n_\text{out} \\ \ell_1 \cdots \ell_m \ell_\text{out}}}^{(m)} = 2\pi\,\delta_{\ell_\text{out},\,\sum_{j=1}^m \text{sign}(\omega_j)\,\ell_j} \int_0^\infty R\, u_{n_1,\ell_1}^{\omega_1}(R) \cdots u_{n_m,\ell_m}^{\omega_m}(R) \times u_{n_\text{out},\ell_\text{out}}^*(R)\, dR,$$

where the superscript in $u_{n_j,\ell_j}^{\omega_j}(R)$ denotes that we should take the complex conjugate of the function for negative frequency components of the field. The Kronecker delta in front of the integral expresses the OAM selection rule as discussed in the main text. Note that, unlike the case of OAM, there is no simple closed form selection rule for the radial index of the eigenmodes $u_{n_{out},\ell_{out}}(R)$. As the interaction between matter and fields in vdW monolayers takes place at the ultimate limit of dimensionality, the coupling and conversion between spatial modes with distinct radial profile is enhanced. Indeed, in the case of bulk materials, the intensity distribution of the electromagnetic waves drifts away from the direction of the wave vectors as the field interacts with matter, a phenomenon known as spatial walk-off. This process deteriorates the overlap between the input pumps and seed beams, thus leading to a decrease in frequency mixing interactions and mode conversion. On the other hand, 2D systems eliminate the possibility of walk-off and propagation-induced mode mismatch. In the case of second order processes with two incident beams (pump and seed), the coupling constant that determines the output idler spatial profile reduces to Eq. (4).

**Acknowledgments:** T.N., L.M.M., L.M.Mc., W.K.K., and P.P. acknowledge support from the Los Alamos National Laboratory - Laboratory Directed Research and Development Program (20220273ER and 20240037DR). L.M.M. also acknowledges support from the Department of Energy (DOE) National Nuclear Security Administration (NNSA) Minority Serving Institution Partnership Program. K.W.C.K. acknowledges support from the DOE NNSA Laboratory Residency Graduate Fellowship program (DE-NA0003960). Work was primarily performed at the Center for Integrated Nanotechnologies, an Office of Science User Facility operated for the U.S. DOE Office of Science. Los Alamos National Laboratory, an affirmative action equal opportunity employer, is managed by Triad National Security, LLC for the US DOE NNSA, under contract no. 89233218CNA000001.

**Author contributions:** The project was conceived by R.P.P. and P.P. Experiments were performed by T.N., L.M.M., L.M.Mc., and N.T. with input from A.J.T., P.J.S., and R.P.P. and under the guidance and supervision of P.P. Theoretical models were developed by W.K.K. with input from J.-X.Z. Crystal growth was performed by J.Y., L.H and J.C.H. Sample preparation was performed by K.W.C.K., N.O., and X.Z. The manuscript was written by T.N., W.K.K., and P.P. with input and final approval from all the coauthors.

**Competing interests:** Authors declare that they have no competing interests.

**Data and materials availability:** All experimental data are available in the main text or the supplementary materials.

**Code availability:** The Mathematica code used to compute the overlap integral and simulated far-field mode intensity distributions will be made available on reasonable request.

**Correspondence and requests for materials** should be addressed to Tenzin Norden, Wilton J. M. Kort-Kamp, and Prashant Padmanabhan.

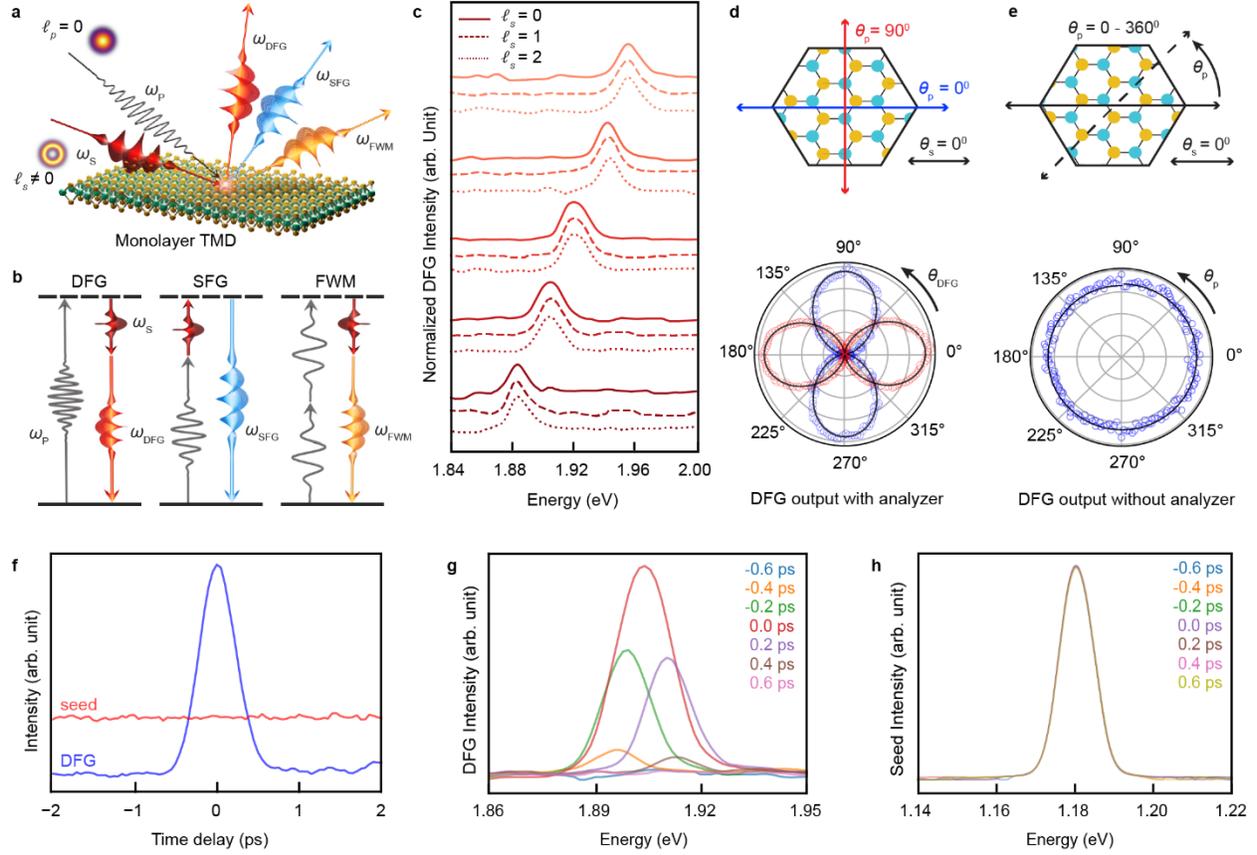

**Figure 1. DFG on monolayer MoS$_2$. a,** Schematic of NLO frequency mixing processes involving a Gaussian pump ($\omega_P$, $\ell_p = 0$) and a vortex seed ($\omega_s$, $\ell_s \neq 0$). **b,** Energy level diagram for the generation of optical vortex DFG ($\omega_{DFG}$), SFG ($\omega_{SFG}$) and FWM ($\omega_{FWM}$) outputs, depicting the conservation of OAM. **c,** Broadband spectrum of DFG output ($\hbar\omega_{DFG} = 1.88 - 1.96$ eV) generated by mixing a Gaussian pump ($\hbar\omega_p = 3.10$ eV) and a vortex seed ($\hbar\omega_s = 1.13 - 1.23$ eV) for $\ell_s = 0$, 1 and 2 (solid, dashed and dotted lines, respectively). **d,** Schematic (top panel) of the polarization angles of the seed ($\theta_s = 0°$, black arrow) and the pump ($\theta_p = 0°$ and 90°, blue and red arrow) with respect to the arm-chair crystallographic axis (i.e., the $x$-axis of the schematic) of a monolayer MoS$_2$. Polar plot (bottom panel) of the DFG output intensity for $\ell_s = 1$ when the seed and the pump polarizations were either collinear ($\theta_p = \theta_s = 0°$, blue pattern) or cross polarized ($\theta_p = 90°, \theta_s = 0°$, red pattern). The curves were fitted with $cos(\theta_{DFG} + \phi)^2$ function, where $\theta_{DFG} = \pi/2 - \theta_s - \theta_p$ and $\varphi$ is an offset angle. **e,** Polar plot (bottom panel) of DFG intensity ($\ell_s = 1$) measured without an analyzer and plotted as a function of the relative polarization angles of the pump and the seed ($\theta_p = 0 - 360°$, $\theta_s = 0°$) as depicted in the schematic (top panel). **f,** Time dependent trace of the DFG output (blue trace) and the reflected seed (red trace) intensity measured by scanning the time delay between the pump and seed pulses (seed trace offset added for clarity). **g,** DFG spectrum as a function of the time delay between the pump and seed pulse. **h,** Seed spectrum as a function of the time delay between the pump and seed pulse.

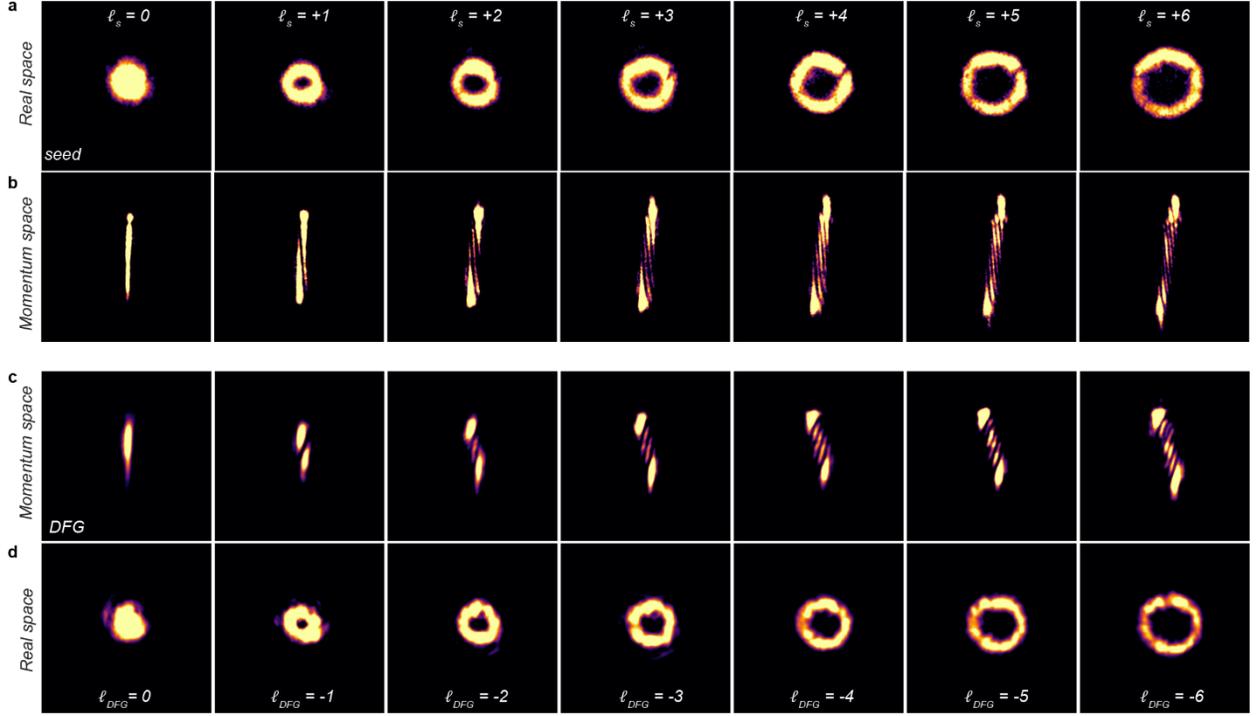

**Figure 2. Intensity profile images of DFG on monolayer MoS$_2$. a,** Real space images of the annular intensity profile of the seed beam ($\hbar\omega_s = 1.18$ eV) for seed topological charges $\ell_s = 0 - 6$. **b,** Momentum space images of the seed at the focal plane of a cylindrical lens (f = 120 mm). The number of fringes, $N_F^S$, is equivalent to $\ell_s + 1$ and the direction of the skew corresponds to the sign of $\ell_s$ (i.e., right to left skew from top to bottom corresponds to positive values). **c,** DFG output ($\hbar\omega_p = 1.92$ eV) momentum space images with number of fringes, $N_F^{DFG} = -N_F^S$, as seen from the opposite skew direction (i.e., left to right from top to bottom), reflecting the fact that $l_{DFG} = -l_s$. **d,** The corresponding real space images of the DFG output. All data were taken with a Gaussian pump ($\omega_p = 3.10$ eV, $\ell_p = 0$). Image intensities were made comparable by adjusting the acquisition settings of the EMCCD.

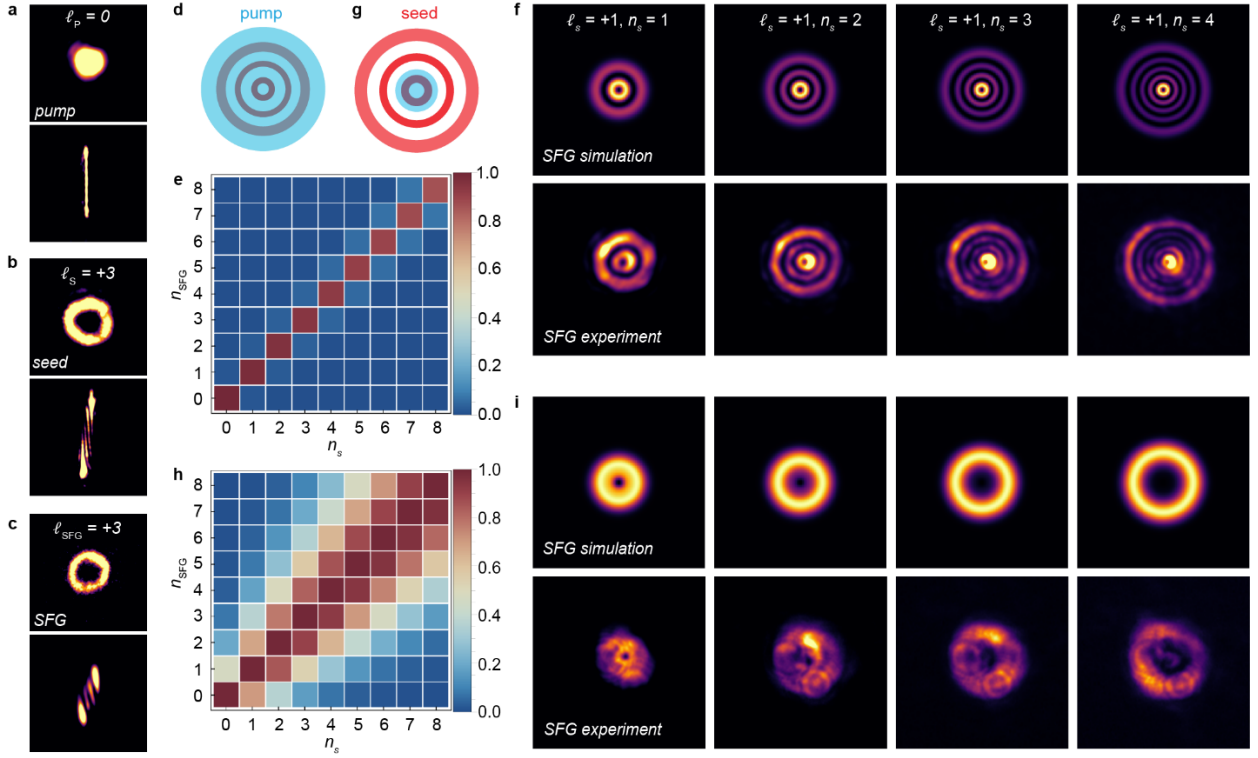

**Figure 3. SFG on monolayer MoS$_2$ with vortex beams. a-c,** Real space (top panel) and momentum space (bottom panel) images of the (a) Gaussian pump ($\hbar\omega_p = 1.63$ eV, $\ell_p = 0$), (b) vortex seed ($\hbar\omega_s = 1.18$ eV, $\ell_s = +3$), and (c) SFG output ($\hbar\omega_{SFG} = 2.81$ eV, $\ell_{SFG} = +3$) where $N_F^S = N_F^{SFG}$. **d,** Illustration of the SFG experimental configuration with a large Gaussian pump beam waist (blue) overlapping the entire vortex seed beam (red) with non-zero radial index, $n_s$. **e,** Overlap integral ($\Lambda$) matrix showing perfect conservation of radial index, $n_{SFG} = n_s$. **f,** Simulations (top panel) and experimentally observed (bottom panel) intensity profiles of SFG output for seeds with $\ell_s = +1$ and $n_s = 1-4$ for the experimental configuration associated with (d) and (e). **g,** Illustration of the SFG experimental configuration where the Gaussian pump beam waist (blue) is same as the seed's (red) central annulus. **h,** Overlap integral ($\Lambda$) matrix showing breakdown of the radial index conservation. **i,** Simulated (top panel) and experimentally observed (bottom panel) SFG output intensity profiles for seeds with $\ell_s = +1$ and $n_s = 1-4$ for the experimental configuration associated with (g) and (h).

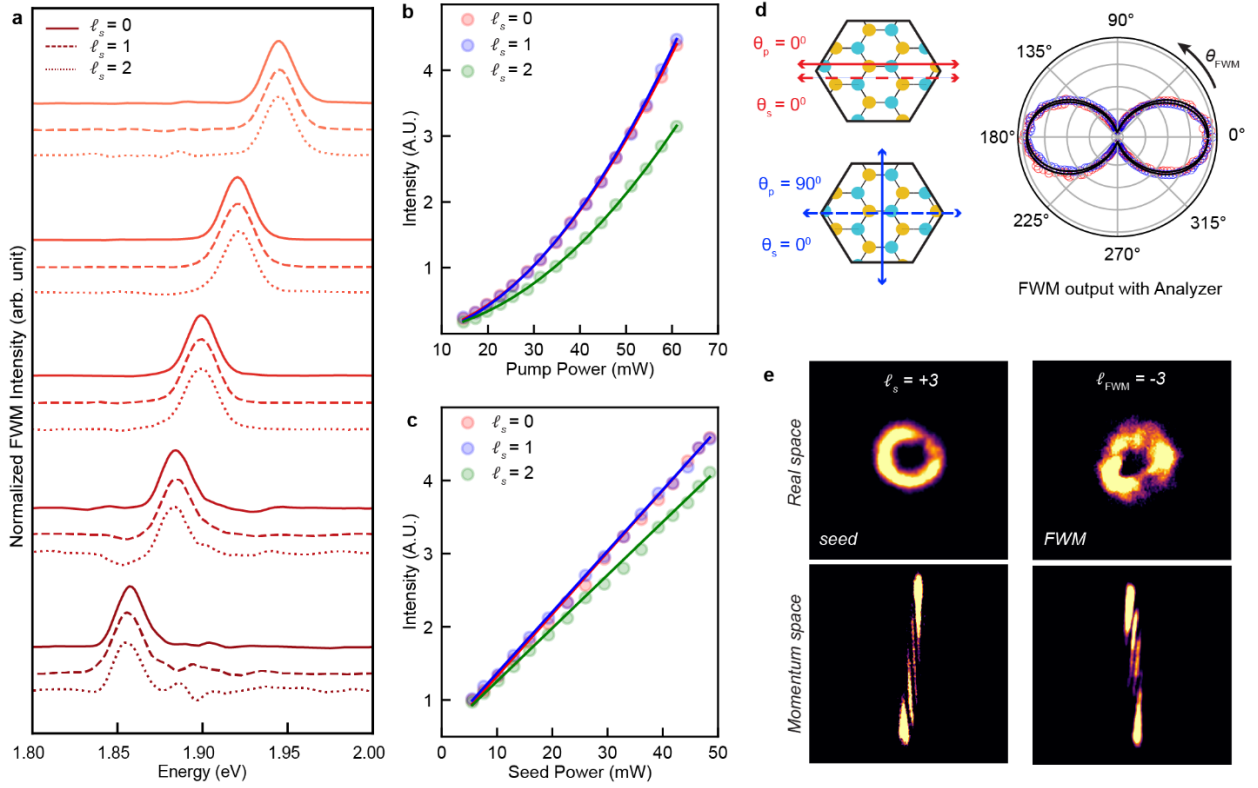

**Figure 4. FWM on monolayer MoS$_2$ with vortex beams. a,** Broad spectrum tuning of the FWM output for seed beams with $\ell_s = 0, 1, 2$ (solid, dashed, and dotted lines, respectively) and Gaussian pump beam ($\hbar\omega_p = 1.54$ eV, $\ell_p = 0$). **b,** Pump power dependence of the FWM output ($\hbar\omega_{FWM} = 1.92$ eV) for $\ell_s = 0, 1, 2$ (red, blue, and green circles, respectively) with solid lines representing square-law fits. **c,** Seed power dependence of the FWM output ($\hbar\omega_{FWM} = 1.92$ eV) for $\ell_s = 0, 1, 2$ (red, blue, and green circles, respectively) with solid lines representing linear fits. **d,** Polar plot (right panel) of the FWM intensity as a function of an analyzer angle ($\theta_{FWM}$) rotated from $0 - 360°$ for $\ell_s = +3$. The schematic (left panel) shows that the seed beam polarization is parallel to the armchair axis of the crystal ($\theta_s = 0$), while the pump beam polarization is either parallel ($\theta_p = 0°$, red circles) or perpendicular ($\theta_p = 90°$, blue circles) to the armchair axis. **e,** Real space annular intensity profile (top panel) and momentum space mapped images (bottom panel) of the seed (left panels) and the FWM output (right panels) for $\ell_s = +3$. The bottom panels show that $N_F^{FWM} = N_F^s$ and patterns with opposite skews.